\pdfoutput=1
\documentclass[a4paper,american,floatfix,pdftex,superscriptaddress,twoside,%
aps,prl,%
citeautoscript,%
reprint
]{revtex4-2}%
\usepackage{amsfonts,amsmath,amssymb}
\usepackage[T1]{fontenc}
\usepackage{graphicx}%
\usepackage[utf8]{inputenc}
\usepackage[todos,ams]{CMImacros}
\usepackage{xr-hyper} 

\graphicspath{{./fig}} 
\hypersetup{citebordercolor=yellow,linkbordercolor=red,urlbordercolor=blue} %

\newcommand{\cfeldesy}{\affiliation{Center for Free-Electron Laser Science CFEL, Deutsches
      Elektronen-Synchrotron DESY, Notkestraße 85, 22607 Hamburg, Germany}}%
\newcommand{\uhhcui}{\affiliation{Center for Ultrafast Imaging, Universität Hamburg, Luruper
      Chaussee 149, 22761 Hamburg, Germany}}%
\newcommand{\uhhphys}{\affiliation{Department of Physics, Universität Hamburg, Luruper Chaussee 149,
      22761 Hamburg, Germany}}%
\newcommand{\stemail}{\email[]{sebastian.trippel@cfel.de}}
\newcommand{\jkemail}{\email[]{jochen.kuepper@cfel.de}}%
\newcommand{\cmiweb}{\homepage[\newline]{https://www.controlled-molecule-imaging.org}}%

\begin{document}
\title{Reaction Pathways of Water Dimer Following Single Ionization}%
\author{Ivo S.~Vinklárek}\cfeldesy%
\author{Hubertus~Bromberger}\cfeldesy%
\author{Nidin~Vadassery}\cfeldesy%
\author{Wuwei~Jin}\cfeldesy%
\author{Jochen~Küpper}\jkemail\cmiweb\cfeldesy\uhhcui\uhhphys
\author{Sebastian~Trippel}\stemail\cfeldesy\uhhcui
\begin{abstract}\noindent%
   Water dimer \HHOd -- a vital component of the earth's atmosphere -- is an important prototypical
   hydrogen-bonded system. It provides direct insight into fundamental chemical and biochemical
   processes, \eg, proton transfer and ionic supramolecular dynamics occurring in astro- and
   atmospheric chemistry. Exploiting a purified molecular beam of water dimer and multi-mass ion
   imaging, we report the simultaneous detection of all generated ion products of
   \HHOdp-fragmentation following single ionization. Detailed information about ion yields and
   reaction energetics of 13 ion-radical pathways, 6 of which are new, of \HHOdp are presented,
   including strong ${}^{18}\text{O}$-isotope effects.
\end{abstract}
\maketitle

\section{Introduction}
\label{sec:introduction}
Water dimer \HHOd is assumed to be a vivid contributor to the radiation
budget~\cite{Chylek:GRL24:2015, Vogt:ARPC73:209}, the homogeneous
condensation~\cite{Schenter:PRL82:3484}, and chemical reactions including degradation of Criegee
intermediates~\cite{Anglada:PCCP18:17698} at low altitudes of the earth's
atmosphere~\cite{Headrick:PCEPC26:479, Tretyakov:PU57:1083,Tretyakov:PRL110:093001}. Moreover, the
decay of the ionic states of water clusters induced by cosmic radiation is considered essential for
astrochemistry occurring on ice mantles~\cite{Molano:AA537:138, Woon:ACR54:490, Ciesla:APJL784:1}
and plays a key role as a trigger of chemical evolution in interstellar
space~\cite{Watanabe:ProgSurfSc83:439}. Therefore, as \HHOd is the smallest water cluster, it is a
favorable model system to develop a more thorough understanding of the environmental effect, \ie,
hydrogen bonding's role, in energy, charge, and mass transfer occurring in ion-radical
chemistry~\cite{Jahnke:NatPhys6:139, Thuermer:NatChem5:590, Jahnke:JPB48:082001,
   Svoboda:PCCP15:11531, Chalabala:JPCA122:3227, Richter:NatComm9:4988, Takada:JPB54:145103,
   Zhang:PRA99:053408}.

Hydrogen bonding is of major importance for the vast majority of biochemical processes and chemical
reactions such as proton transfer in redox reactions of light-harvesting
complexes~\cite{Kramer:TPS9:349}, or structure and topological stability of bio\-mole\-cules like
the double-helix structure of DNA~\cite{Boudaiffa:Science287:1658, Hanel:PRL90:188104}.
Specifically, understanding how absorbed energy dissipates and charge is re-distributed within an
ionized aqueous environment is crucial in a biological context, where the interaction of
biomolecules with generated low-energy electrons and ion-radicals can result in their degradation.

The interest in ionic supramolecular dynamics of the \HHOd system led to several theoretical
predictions~\cite{Svoboda:PCCP15:11531, Chalabala:JPCA122:3227, Takada:JPB54:145103,
   Svoboda:JCP135:154302} of fragmentation pathways of singly and doubly charged water dimer and
fragmentation energetics. These were only partly supported by the experimental results of
electron-impact ionisation~\cite{Buck:ZP31:291} and of Coulomb explosion~\cite{Jahnke:JPB48:082001,
   Jahnke:NatPhys6:139, Chng:JACS134:15430, Zhang:PRA99:053408, Schnorr:SciAdv9:7864} induced by
core and valence ionization. This is not surprising, considering the experimental challenge, \eg,
the interference of the products of \HHOdp fragmentation with those resulting from the fragmentation
of higher clusters and isolated \HHO molecules. Our experimental approach circumvents these issues
by using the electrostatic deflector~\cite{Bieker:JPCA123:7486} for the purification of \HHOd
samples and multi-mass ion imaging~\cite{Bromberger:JPB55:144001, Cheng:RSI93:013003,
   Brouard:RSI83:114101} to monitor all the ion-radical channels induced by strong-field ionization
at once. The photoionization was set to be predominantly in the multiphoton regime to populate
similar states of the \HHOdp ion as those populated from VUV/UV- or electron impact ionization.

\textit{Ab initio} simulations revealed that the photoionization of \HHOd through electron ejection
from the four highest occupied molecular orbitals (HOMO to HOMO$-$3) preferably leads to proton
transfer~\cite{Svoboda:PCCP15:11531}. The first two ionic states of \HHOdp, \ie, the ground
${}^2A^{''}$ and the first excited ${}^2A^{'}$ states, are populated by electron ejection from the
two non-bonding $\text{1b}_1$ orbitals of oxygen in the proton donor (HOMO) and the proton acceptor
(HOMO$-$1), respectively~\cite{Svoboda:PCCP15:11531, Takada:JPB54:145103}. Both of these states are
highly reactive, which is underlined by the estimated timescale for proton migration of less than
100~fs and 300~fs for \HHOdp in its ${}^2A^{''}$ and ${}^2A^{'}$ states,
respectively~\cite{Svoboda:PCCP15:11531}. Recently, the timescale for the proton migration was
measured as 55(20)~fs in XUV-pump-XUV-probe experiments~\cite{Schnorr:SciAdv9:7864}. Subsequently,
\HHOdp either dissociates into
\begin{equation}
   \label{eq:H3O++OH}
   \HHOdp \rightarrow  \HHHOp  + \text{OH},\enspace E_{\text{A}} =
   11.7\, \text{eV},
\end{equation}
or survives as an ion-radical pair ($<40\,\%$)~\cite{Chalabala:JPCA122:3227, Svoboda:PCCP15:11531}
\begin{equation}
   \label{eq:(H2O)2+}
   \HHOdp \rightarrow (\HHHOp\text{-OH}),\enspace E_{\text{A}} =
   11.7\, \text{eV}.
\end{equation}
$E_{\text{A}}$ denotes the specific appearance energy, \ie, the minimum energy required to induce
the reaction pathway.

The next two excited states of \HHOdp are populated through electron ejection from the $\text{3a}_1$
orbital of donor (HOMO$-$2) and acceptor (HOMO$-$3) \HHO followed by the prompt
decay~\cite{Svoboda:PCCP15:11531} into the ${}^2A^{''}$ and ${}^2A^{'}$ states with subsequent
dissociation \emph{via} the channel in \eqref{eq:H3O++OH}. Additional minor channels ($<20~\%$)
either lead to an ion radical pair \HHOdp through the channel in \eqref{eq:(H2O)2+} or fragmentation
without proton transfer as
\begin{equation}
   \label{eq:H2O++H2O}
   \HHOdp \rightarrow \HHOp + \HHO, \enspace E_{\text{A}} = 12.8\, \text{eV}.
\end{equation}
Furthermore, the electron ejection from the $\text{1b}_2$ orbital of the donor (HOMO$-$4) and acceptor
(HOMO$-$5) predominantly leads to the three-body fragmentation as
\begin{equation}
  \label{eq:H2O++OH+H}
  \HHOdp \rightarrow \HHOp + \text{OH} + \text{H},\enspace E_{\text{A}} = 18.2\,
  \text{eV},
\end{equation}
generating four species: three radical fragments and an electron. Those species should eminently
increase the possibility of radiation damage in an aqueous environment~\cite{Svoboda:PCCP15:11531}.
The two-body fragmentation channels producing \HHHOp and \HHOp ions participate only with a minor
contribution ($<20\,\%$) after HOMO$-$4 and HOMO$-$5 ionization~\cite{Svoboda:PCCP15:11531}.

Additional channels of \HHOdp fragmentation appearing at higher ionization energies were
described~\cite{Svoboda:PCCP15:11531,Chalabala:JPCA122:3227}, \ie,
\begin{alignat}{3}
    \HHOdp &\rightarrow \text{OH}^{+} + \HHO + \text{H},&& \ E_{\text{A}} = 18.5\, \text{eV}, \label{eq:OH++H2O+H} \\
    \HHOdp &\rightarrow \text{H}^{+} + \HHO + \text{OH},&& \ E_{\text{A}} = 19.2\, \text{eV}, \label{eq:H++H2O+OH} \\
    \HHOdp &\rightarrow (\HHO\text{-O})^{+} + 2\text{H},&& \ E_{\text{A}} = 32.6\, \text{eV}. \label{eq:H2O-O++2H}
\end{alignat}

Contrary to valence ionization, stripping of a core electron from the $\text{2a}^1$ or $\text{1a}^1$
orbitals, \eg, by x-ray ionization, induces direct or
proton-transfer-mediated~\cite{Richter:NatComm9:4988, Thuermer:NatChem5:590} relaxation through
local or intermolecular Auger decay~\cite{Jahnke:NatPhys6:139,Jahnke:JPB48:082001,
   Richter:NatComm9:4988, Thuermer:NatChem5:590} and subsequent ejection of a valence electron.

Here, we exploited the combination of electrostatic deflection~\cite{Chang:IRPC34:557} and
multi-mass imaging with a Timepix3 camera~\cite{Bromberger:JPB55:144001, Cheng:RSI93:013003} to
yield unprecedented details into the ionic reaction pathways of the prototypical ionized \HHOd
system. We produced a rotationally cold molecular beam of \HHOd with $\ordsim92~\%$
purity~\cite{Bieker:JPCA123:7486} and recorded all ionic reaction products, see Methods and the
supporting information (SI) for details. We directly observed all theoretically predicted
fragmentation channels of \HHOdp~\cite{Svoboda:PCCP15:11531, Chalabala:JPCA122:3227}. Moreover, we
observed multiple new fragmentation channels. The velocity-map-imaging (VMI) detection inherently
provided information about the released translational energy and the rovibronic excitation of the
products, which could affect subsequent reactions.

\section{Methods}
\label{sec:setup}
\begin{figure}[b]
   \includegraphics[width=\linewidth]{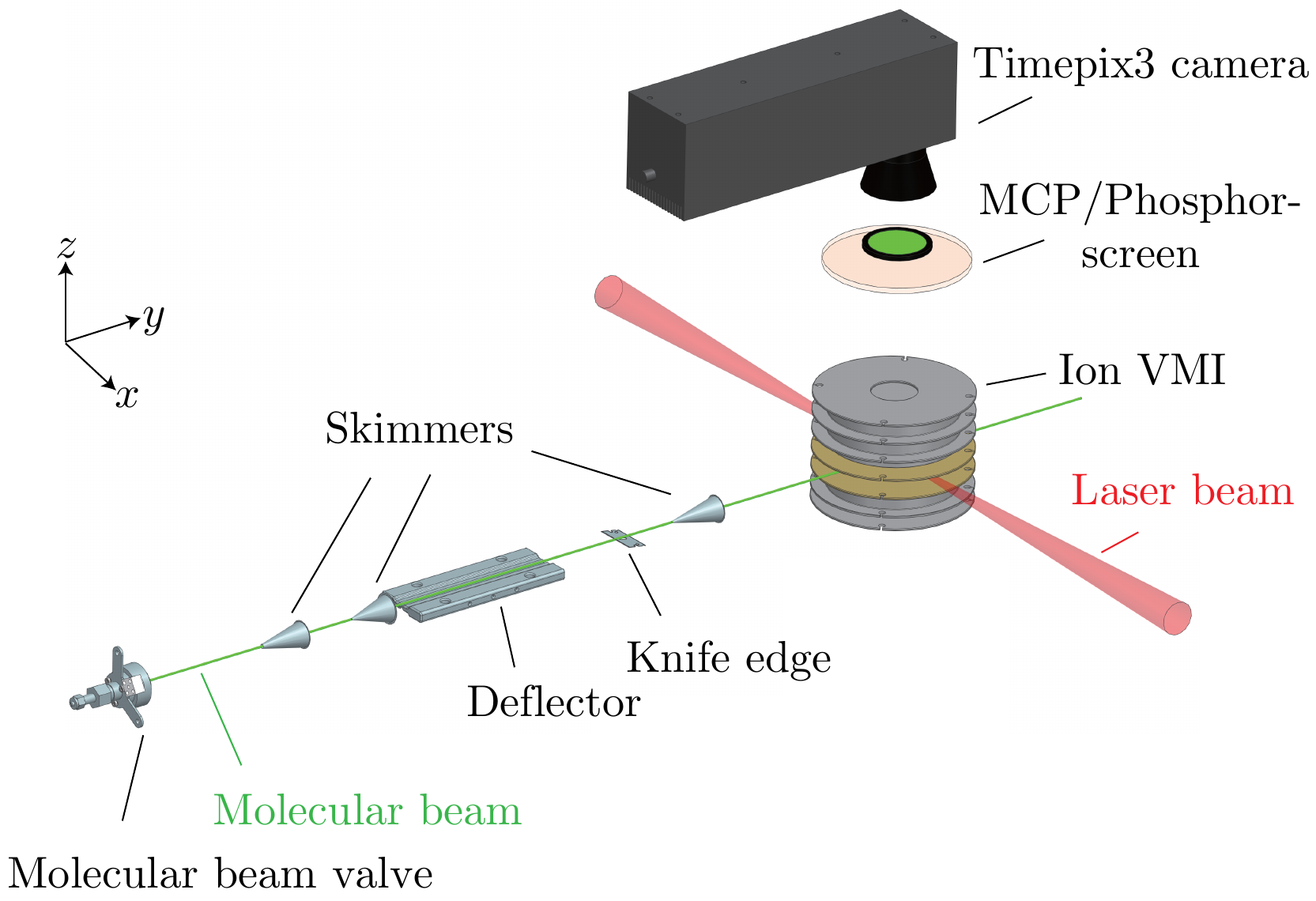}
   \caption{Schematic picture of the endstation for controlled-molecule experiments
      (eCOMO)~\cite{Trippel:RSI89:096110, Jin:eCOMO:inprep}.}
   \label{fig:experimental_setup}
\end{figure}

All experiments were performed in our recently commissioned transportable endstation for
controlled-molecule experiments (eCOMO), which will be described in more detail
elsewhere~\cite{Jin:eCOMO:inprep}. \autoref{fig:experimental_setup} provides a sketch of the
experimental setup~\cite{Jin:eCOMO:inprep, Trippel:RSI89:096110}.
A sample of distilled water (0.5~\ulit, room temperature) was dropped on a glass-filter paper and
installed in a sample holder behind the pulsed valve (Amsterdam
Piezovalve)~\cite{Irimia:RSI80:113303}. The water vapor was mixed with helium buffer gas using a
stagnation pressure of 4~bar and the mixture coexpanded into vacuum in 50~\us pulses into the source
chamber; pressures with the valve on/off were $2\times10^{-6}~\rm{mbar}/2\times10^{-8}~\rm{mbar}$.
The molecular beam of water clusters was extracted by a first skimmer (diameter 3~mm) from the
supersonic jet and further collimated by a second skimmer (diameter 1.5~mm) before entering the
electrostatic deflector.

To spatially separate the water dimer clusters from the carrier helium gas as well as from isolated
water molecules and larger water clusters, we used a $b$-type deflector~\cite{Chang:IRPC34:557,
   Kienitz:JCP147:024304}, applying a voltage of 13~kV. The dipole moments of water monomer and
dimer are 1.86~D and 2.63~D,
respectively~\cite{Gregory:CPL282:147,Bieker:JPCA123:7486,Gregory:Science275:814}. Subsequently, the
molecular beam was cut in half by a knife edge to increase both, the effective separation of
water-dimer from the rest of the molecular beam and the column density~\cite{Trippel:RSI89:096110}.
The source, deflector, knife edge, and skimmers are all movable, which enabled us to optimize their
positions for the separation and purification of the water-dimer clusters. The deflected molecules
were intersected with short laser pulses in the interaction region of a double-sided
velocity-map-imaging (VMI) spectrometer. The purity of the water-dimer beam was estimated to be
92~\% by comparing the signals from known fragmentation channels compared to the background signal,
\ie, water monomer, nitrogen, and oxygen, at the deflected-beam position of 2~mm.

To photoionize the molecules, we used a 800~nm Ti:Sapphire chirped-pulse-amplifier system (Coherent
Astrella) operated at 1~kHz. The pulse duration is estimated to 40~fs and the pulse
energies of 170~\uJ. Focusing the laser beam to 53~\um full width at half
maximum (FWHM) intensity yielded a peak intensity up to $\sim2\times10^{14}~\rm{W}/\rm{cm}^2$,
corresponding to a Keldysh parameter of $\sim0.7$ for photoionization of \HHOd at
$\Ei=11.7$~eV~\cite{Snow:IJMS96:49}.

The generated ions were then accelerated toward the detector in a perpendicular-geometry of the
double-sided spectrometer~\cite{Eppink:RSI68:3477} and projected under VMI conditions. At the end of
the spectrometer, a micro-channel-plate phosphor-screen combination (MCP: Photonics, APD 3 PS
75/32/25/8 I 60:1 NR MGO 8''FM; PS: P47) was mounted to produce light flashes for individual ions.
The flashes were detected with a Timepix3 camera~\cite{Bromberger:JPB55:144001, Zhao:RSI88:113104}
(Amsterdam Scientific Instruments) operated and controlled by our open-source-library
Pymepix~\cite{AlRefaie:JInst14:P10003, pymepix:github}, which was also used to extract the raw
physics events from the Timepix3 data stream. The detectors temporal resolution of
$\ordsim1.6$~ns~\cite{Bromberger:JPB55:144001} enabled us to work in multi-mass-detection mode and
to obtain the VMI images of all fragments directly by slicing, \ie, computational event selection,
in the time-of-flight coordinate. The synchronization of the whole experiment is provided by a
Stanford DG645 delay generator.

The plotted VMI images and beam-plot figures were background subtracted and also stripped off the
contribution from the scattered/non-deflected beam by subtraction of the measured
background-subtracted signal with the deflector off. The valve was operated at 200~Hz resulting in 4
successive background measurements due to the 5 times higher repetition rate of the laser with
respect to the molecular-beam valve. The data were centroided by calculating the center-of-mass for
each detected ion~\cite{Bromberger:JPB55:144001}. The 3D velocity (radial) and angular distributions
were obtained by standard integration methods for VMI images~\cite{Whitaker:Imaging:2003}. For
momentum calibration~\cite{Bromberger:JPB55:144001}, we fitted the dependence of the center
positions of the VMI images on the time-of-arrival and converted it to momenta applying the physical
size of the ion detector and the magnification factor ($M = v_\text{i}/v_\text{mb} = 0.843$) of the
spectrometer estimated by SIMION~\cite{Simion:8.1}. The resulting velocity of the molecular beam
is $2~\text{km/s}$.

\section{Results and Discussion}
\label{sec:results}
\begin{figure*}
   \includegraphics[width=\linewidth]{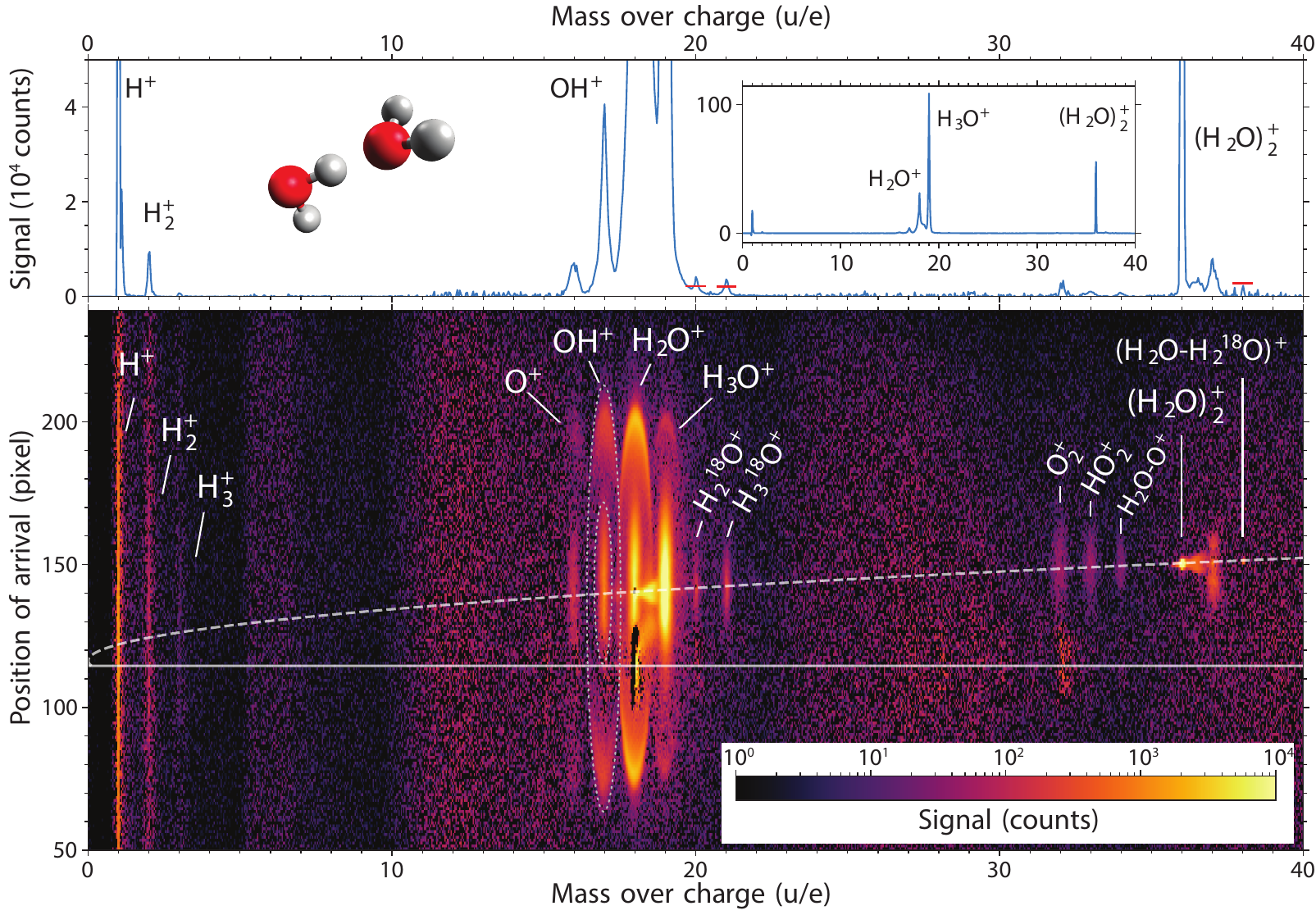}%
   \caption{Image mapping the ion signal according to the mass-over-charge ratio and the position of
      arrival on the detector (lower panel). The upper panel outlines the zoomed-in mass spectrum,
      with an inset showing the whole vertical range for reference. The observed fragments of \HHOdp
      are $\text{H}^+$, $\text{H}_2^+$, $\text{H}_3^+$, $\text{O}^+$, \OHp, \HHOp, \HHHOp,
      $\text{O}_2^+$, $\text{HO}_2^+$ and ($\HHO\text{-O}$)$^{+}$. The observed fragments of
      $(\HHO\text{-H}_2{}^{18}\text{O})^+$ are $\text{H}_2{}^{18}\text{O}^+$ and
      $\text{H}_3{}^{18}\text{O}^+$. The expected isotopologue-peak heights of
      $\text{H}_2{}^{18}\text{O}^+$, $\text{H}_3{}^{18}\text{O}^+$, and
      $(\HHO\text{-H}_2{}^{18}\text{O})^+$ ions are indicated by the red horizontal bars. Dotted
      ellipses at \mq{17} are drawn to mark product regions of \HHOdp fragmentation and Coulomb
      explosion as an example. See text for details.}
   \label{fig:beamplot}
\end{figure*}
\autoref{fig:beamplot} depicts an overview of all acquired ions and their fragmentation channels
resulting from singly and multiple-charged water dimer after strong-field ionization at $800$~nm.
The image shows the background subtracted ion signal for the mass-over-charge ratio and the position
of arrival at the detector, see SI for details. The mass-over-charge ratio $m/q$ is obtained by a
direct transformation from the ion time of flight by $m/q\propto t^2$. Due to the molecular beam
velocity, in this two-dimensional (2D) spectrum, all signals from species originating from the
molecular-beam are shifted upward in the figure with increasing mass-over-charge ratio. Ions with
zero velocity in the co-moving frame of the molecular beam appear on the dashed line. Ions in the
surrounding structures centered at the dashed line are due to fragmentation of the cluster after
ionization. Remaining signals are due to the ionization of the diffuse rest gas in the chamber.

For a direct comparison between the various fragments in terms of signal strength, the inset in the
top graph of \autoref{fig:beamplot} depicts the mass spectrum obtained by summation of the signal
in the lower figure along the vertical axis. The top figure itself is a zoom into the inset to
highlight weak channels.

\begin{table*}
   \centering
   \caption{Summary of all the observed ions from \HHOdp dissociation together with their relative
      ion yields. Our estimated appearance energies $E_{\text{A}}$ are shown in the fourth column in bold,
      alongside the previously reported appearance energies~\cite{Svoboda:PCCP15:11531,
         Chalabala:JPCA122:3227}. The last column shows previously reported and suggested
      fragmentation pathways producing the observed ions.}%
   \label{tab:branchingratios}
   \begin{tabular}{lrlcl}
     \toprule
     Fragment  & \multicolumn{1}{l}{Signal} & \multicolumn{1}{l}{Relative} & $E_{\text{A}}$~(eV) & Reaction channel \\
               & \multicolumn{1}{l}{(counts)} & \multicolumn{1}{l}{ion yield} & &  $\HHOdp \rightarrow $\\
     \hline
     \HHHOp & 1933751 & $0.541(42)$ & 11.7~\cite{Svoboda:PCCP15:11531} & $\HHHOp + \text{OH}$~\cite{Svoboda:PCCP15:11531, Chalabala:JPCA122:3227, Zhang:PRA99:053408} \\
     \HHOp & 630492  & $0.176(11)$  & \begin{tabular}{c}12.8~\cite{Svoboda:PCCP15:11531}\\  18.2~\cite{Svoboda:PCCP15:11531} \end{tabular}  & \begin{tabular}{l} $\HHOp + \HHO$~\cite{Svoboda:PCCP15:11531, Chalabala:JPCA122:3227, Zhang:PRA99:053408} \\  $\HHOp + \text{OH} + \text{H}$~\cite{Svoboda:PCCP15:11531} \end{tabular} \\
     \HHOdp & 613728 & $0.172(8)$ & 11.7~\cite{Svoboda:PCCP15:11531} & $\HHHOp\cdots \text{OH}$~\cite{Svoboda:PCCP15:11531, Chalabala:JPCA122:3227, Zhang:PRA99:053408} \\
     $\text{H}^+$ & \textsuperscript{[a]}213905 & $0.06(4)$ & 19.2~\cite{Svoboda:PCCP15:11531} & $\text{H}^{+} + \text{H}_2\text{O} + \text{OH}$\cite{Svoboda:PCCP15:11531} \\
     \OHp & 114390 & $0.032(8)$ & 18.5~\cite{Svoboda:PCCP15:11531} & $\text{OH}^{+} + \HHO + \text{H}$~\cite{Svoboda:PCCP15:11531} \\
     $\text{O}^+$ & 24235 & $0.0068(20)$ & \textbf{21.7} & $\text{O}^{+} + \HHO + 2\text{H}$ \\
     $\text{H}_2^+$ & 13947 & $0.0068(14)$ & \textbf{21.7} & $\text{H}_2^{+} + \HHO + \text{O}$ \\
     $\text{H}_2{}^{18}\text{O}^+$ & 10225 & $0.0029(1)$ & - & - \\
     $\text{H}_3{}^{18}\text{O}^+$ & 9268 & $0.0026(3)$ & - & - \\
     $\text{O}_2^+$ & 6052  & $0.0017(3)$ & \textbf{27.6} & $\text{O}_2^{+} + 4\text{H}$ \\
     $\text{HO}_2^+$ & 4206 & $0.0012(2)$ & \textbf{29.2} & $\text{HO}_2^{+} + 3\text{H}$ \\
     $(\HHO\text{-O})^+$ & 1792 & $0.0005(1)$ & 32.6~\cite{Chalabala:JPCA122:3227}/\textbf{33.0} & $(\HHO\text{-O})^{+} + 2\text{H}$~\cite{Chalabala:JPCA122:3227} \\
     $\text{H}_3^+$ & 796  & $0.0002(3)$ & \textbf{36.5} & $\text{H}_3^{+} + \text{HO}_2$ \\
     \hline
   \end{tabular}
   \linebreak%
   \footnotesize{\textsf{[a] Signal contributions by fast non-detected $\text{H}^+$ ions are
         expected to be below 5~\% of the integrated $\text{H}^+$ signal.}}
\end{table*}

The most prominent features in \autoref{fig:beamplot} are observed in the region between $m/q=16$ to
$19$~u/e, corresponding to the $\text{O}^+$, \OHp, \HHOp and \HHHOp fragments. These fragments
originated from the \HHOdp fragmentation and Coulomb-explosion channels of \HHOdpp. We were able to
distinguish between these two sources of the signal using the 3D ion velocity detection with the
Timepix3 camera~\cite{Bromberger:JPB55:144001}. Whereas \HHOdp fragmentation is represented by the
central broad features around the dashed line, \eg, confined by a inner dotted ellipse at $m/q=17$,
the Coulomb explosion channels from \HHOdpp exhibit sharp ring-like -- graphically-projected oval --
structures assigned to ``fast'' ions, \eg, the area between the two dotted ellipses around $m/q=17$,
which will be discussed in a future publication. The strong peaks at $m/q=16$ to $19$~u/e are also
accompanied by weak signals at~$m/q=20$ and $21$~u/e assigned to the isotopologues
$\text{H}_2{}^{18}\text{O}^+$ and $\text{H}_3{}^{18}\text{O}^+$, respectively. The signals
corresponding to the expected peak heights for isotopes in natural abundance, \ie, $0.2$~\% of the
$^{16}\text{O}$-isotopologue signal, are indicated by the red horizontal lines in the mass spectrum
of \autoref{fig:beamplot}. The red bar assigned to $\text{H}_2{}^{18}\text{O}^+$ was corrected
upward to compensate for the contribution from the neighboring \HHHOp signal.

The second region with a strong signal, at $m/q=36$~u/e, corresponds to the parent ion \HHOdp.
Moreover, it is accompanied by a weak peak of its isotopologue
$(\HHO\text{-H}_2{}^{18}\text{O})^{+}$ at $m/q = 38 $~u/e, again with its expected natural-abundance
signal contribution indicated by a red line in the mass spectrum. The signal strength is in very
good agreement with the expected abundance. The well-resolved and clear observation of these
isotopologue demonstrates the high sensitivity of our experiment. In between the two peaks there is
another structure at $m/q=37$~u/e, which we assign to protonated water dimer $\HHOd\text{H}^+$
originating from the fragmentation of larger clusters $(\text{H}_2\text{O})_n^{+}$, $n > 2$, \ie, a
remaining impurity in the experiment.

For $m/q$ ratios of $32$, $33$, and $34$~u/e we also observed weak but distinct signals originating
in the molecular beam. These signals were assigned to $\text{O}_2^+$, $\text{HO}_2^{+}$ and
$(\HHO\text{-O})^{+}$ ions~\cite{Chalabala:JPCA122:3227}, respectively. The origin of these
fragments is not straightforwardly linked to any specific ion precursor. Nevertheless, measurements
without purification of the reactant by deflection do not exhibit any significantly intense signals
between $m/q=32$ and $34$~u/e, see \autoref{esi-fig:beamplot-non-deflected} in SI, which excludes
their origin from larger water clusters $(\text{H}_2\text{O})_n$ or mixed
$(\text{H}_2\text{O})_m(\text{O}_2)_n$ clusters. Furthermore, there are no coincidences of the
$\text{O}_2^{+}$, $\text{HO}_2^{+}$, and $(\HHO\text{-O})^{+}$ ions with $\text{H}^+$ and
$\text{H}_2^+$ ions, see \autoref{esi-fig:pipico0to40} in SI. Therefore, we assign the signal at
$32$, $33$, and $34$~u/e to fragmentation products of \HHOdp.

The last region with a distinguishable signal is located at $m/q = 1$--$3$~u/e. We assign the signal
to the hydrogen ions $\text{H}^+$, $\text{H}_2^+$ and $\text{H}_3^+$ as the isotopic
natural-abundance contribution of deuterium $\text{D}$ is below~$0.02~\%$.

\begin{figure}[t]
   \includegraphics[width=\columnwidth]{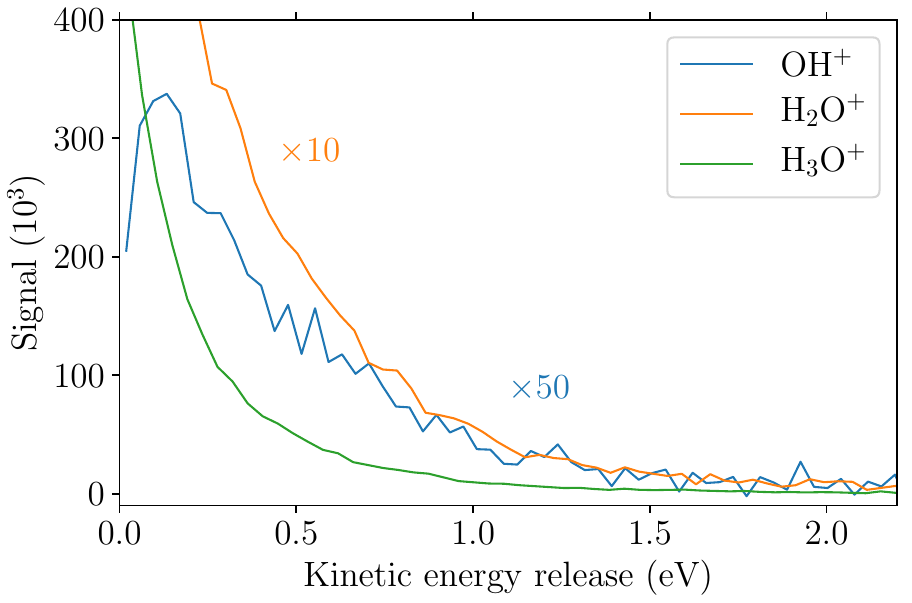}%
   \caption{Kinetic energy release of the reaction channels yielding the specified ions,
      assuming two-body fragmentation.}
   \label{fig:TotalEnergy}
\end{figure}
Besides the mass spectrometric information about the created ions and their signal, in VMI we also
obtain insight into the energy redistribution for the specific fragmentation channels through the
distributions of kinetic energy releases, which are displayed for some selected ions in
\autoref{fig:TotalEnergy}; for comparison, see also the released-total-momentum distributions in
\autoref{esi-fig:TotalMomentum} of the SI. The kinetic-energy-release spectra were obtained from the
measured 3D single ion momenta assuming two-body fragmentation of the parent ion.

This is a valid assumption for the \HHHOp ions and the larger part of the \HHOp signal according to
channels~\eqref{eq:H3O++OH} and \eqref{eq:H2O++H2O}, respectively, with the calculated
kinetic-energy release equal to the total kinetic energy $E_{\text{tr}}$. However, the \OHp ions
were produced from three-body fragmentation according to channel~\eqref{eq:OH++H2O+H}. Here, the
most probable scenario is a sequential fragmentation into \OHp with an intermediate
$\text{H}_3\text{O}$, which further dissociates into \HHO and H. Thus, the plotted curve of the
kinetic energy release in the \OHp reaction channel illustrates the kinetic energy released in the
first step in the three-body fragmentation.

The low-energy peaks up to 2~eV in the $E_{\text{tr}}$ spectra are attributed to the decay of
various rovibrational states of \HHOdp. The individual curve shape then results from the convolution
of initial rovibronic excitation of \HHOdp and subsequent statistical fragmentation. From these
distributions, rovibronic excitation energies $E_{\text{rv}}$ can be obtained assuming the
conservation of energy given by $E_{\text{total}} = E_{\text{tr}} + E_{\text{rv}}$. In a
state-selective experimental approach, such a constraint would allow for the calculation of the
exact rovibronic excitation of the generated fragments. However, limitations arise in the
ultrashort-pulse-ionization regime due to the spectral bandwidth of the ionizing laser,
$\ordsim30$~nm in our case. Assuming that the deposited energy during the \HHOd ionization is given
by the lowest number of photons necessary to overcome the energy threshold, \ie, the appearance
energy $E_\text{A}$, of the selected fragmentation-reaction channel and considering the effective
spectral narrowing due to the multiphoton ionization, one could extract low-resolution
($\larger0.04$~eV) rovibrational spectra using the dissociation threshold for each detected
fragment.

\autoref{tab:branchingratios} provides a summary of previously reported- and our new findings: A
list of all detected ions from decay of \HHOdp is shown in the first column. Seven of the observed
ions can be linked to the already reported pathways from simulations~\cite{Chalabala:JPCA122:3227,
   Svoboda:PCCP15:11531}. The additionally detected ions are created through previously not reported
fragmentation channels. The fragmentation pathways leading to the ejection of the newly discovered
ions and those discussed in the introduction are shown in the last column. These suggested pathways
are based solely on energetic arguments and thus should be further investigated, \eg, by
computations or through radical detection. From our data, we cannot directly link the detected ions
to the exact initial state of the ion. Nevertheless, we can count the number of detected ions and
determine their relative ion yields, which are listed in the second and third columns, respectively.

\begin{figure}[b]
   \includegraphics[width=\columnwidth]{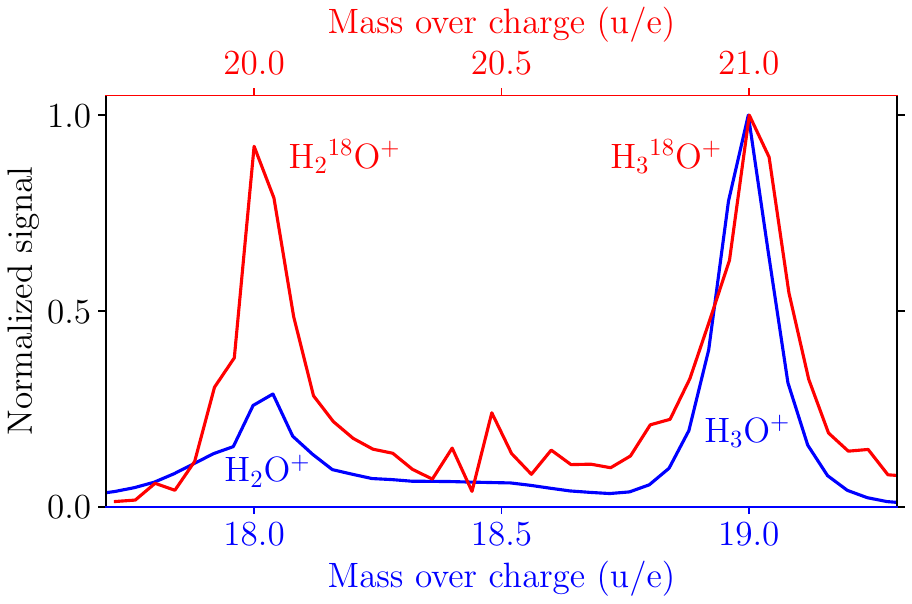}%
   \caption{Comparison of the peak-normalized mass spectra including the signals assigned to (blue)
      \HHOp and \HHHOp, and (red) $\text{H}_2{}^{18}\text{O}^+$ and $\text{H}_3{}^{18}\text{O}^+$.
      The red line is background corrected to exclude the background of the
      $\text{H}_2{}^{18}\text{O}^+$ peak centered at 20~u/e. This is shown in
      \autoref{esi-fig:isocompMS} in SI. See also the text for details on possible small
      ${}^{17}\text{O}$ backgrounds.}
   \label{fig:MSisotop}
\end{figure}
Inspecting the ion yields, an intriguing observation is that whereas the signal intensity of \HHHOp
is roughly three times larger than the signal of \HHOp, the situation is changed for their
isotopologues $\text{H}_3{}^{18}\text{O}^+$ and $\text{H}_2{}^{18}\text{O}^+$, which exhibit
comparable signal intensities, see \autoref{fig:MSisotop}. With the natural abundance of
${}^{17}\text{O}$ and the sensitivity of our experiment, one would also expect to observe
$\text{H}_3{}^{17}\text{O}^+$ and $\text{H}_2{}^{17}\text{O}^+$. Unfortunately, these signals are
hidden under the stronger signals of $\text{H}_2{}^{18}\text{O}^+$ and \HHHOp, respectively. For the
\HHHOp signal, a background contribution from $\text{H}_3{}^{17}\text{O}^+$ is below 0.04~\% and
thus negligible. For the $\text{H}_2{}^{18}\text{O}^+$ peak at \mq{20}, a contribution by
$\text{H}_3{}^{17}\text{O}^+$ could be as large as 20~\%. In any case, the difference of the ratios
for the $\text{H}_{2,3}{}^{16}\text{O}$ and $\text{H}_{2,3}{}^{18}\text{O}$ signals is a
surprisingly strong illustration of isotopic substitution modifying chemical-reaction pathways. This
could be due to a preferred stereometric position of $\text{H}_2{}^{18}\text{O}$ as a proton donor
in \HHOd, or it could be due to a reduced proton tunneling and transfer probability between the two
water moieties in the ${}^{18}\text{O}$ isotopologue, which could both be a consequence of small
reduced-mass changes resulting in zero-point-energy and anharmonic-coupling effects.

On top of the qualitative analysis, the obtained relative ion yields were also utilized to estimate
the appearance energies $E_{\text{A}}$ of the newly detected ions and assigned channels. Within our
model, the relative ion yields can be expressed as
\begin{equation}
   \label{eq:ion_yield}
   P(\text{I}^+) = N\cdot\sum_{k}y^{S}_{k}(\text{I}^+)\cdot D(F, \Ei(S))
\end{equation}
$N$ is a normalization factor and $y^{S}_k$ is the branching ratio of the channel $k$ producing ion
$\text{I}^+$ after ionization into an initial ionic state $S$. The probability to ionize into such a
state $S$ is described by the distribution $D$ as a function of the ionization energy \Ei of the
state $S$ and a parameter $F$ corresponding to the strength of the applied external electric
field~\cite{Ammosov:SVJETP64:1191}. The $D$ was chosen to have an exponential character based on the
strong-field ionization approximation~\cite{Ammosov:SVJETP64:1191}. The unknown parameters of our
model, \ie, $F$ and the branching ratios of $\text{H}^+$ and \OHp ions, were calibrated using the
theoretical branching ratios and $E_{\text{A}}$ of each channel given by
Svoboda~\etal~\cite{Svoboda:PCCP15:11531}. We then calculated the so far unknown $E_{\text{A}}$ of
the less abundant ions from \eqref{eq:ion_yield}, see the SI for further details. These estimated
$E_{\text{A}}$ are shown in \autoref{tab:branchingratios} including the reported
ones~\cite{Svoboda:PCCP15:11531, Chalabala:JPCA122:3227}. This includes the $E_{\text{A}} = 33.0$~eV
of $(\HHO\text{-O})^+$ known from~\cite{Chalabala:JPCA122:3227} to be $32.6$~eV. The surprisingly
good agreement should be taken with caution and not as an illustration of the high precision of the
presented \textit{ad hoc} model. The calculated and reported appearance energies $E_\text{A}$
assigned to the observed ions directly reflect the broad range of the initial energies deposited by
the strong-field photoionization triggering the \HHOdp-fragmentation reactions.

\section{Conclusions}
\label{sec:conclusions}
Overall, our study provides unique novel experimental observations of the \HHOdp fragmentation
pathways, which were previously only, and only partly, predicted by molecular dynamics
simulations~\cite{Svoboda:PCCP15:11531, Chalabala:JPCA122:3227}. These results substantially broaden
our perspective on \HHOdp fragmentation by showing an additional set of newly observed ion-radical
pathways and their relative ion yields. Together with our estimated appearance energies
$E_{\text{A}}$, these indicate the relative significance among all detected channels.

The observed very strong O-isotope effect indicates non-classical aspects of molecular
structure and reactivity in the fragmentation. Together with the experimental information on kinetic
energies and angular distributions of all fragments, these data is a very valuable asset for
advanced molecular-dynamics simulations unraveling the underlying chemistry. Time-resolved
investigations of the ionization and subsequent reaction pathways of water dimer could provide
further insight into its electronic and nuclear dynamics, including the timescales of proton and
hydrogen transfer, the traversal of electronic states, and electronic relaxation processes.

Our findings are relevant for discussions of the role of water ionization and the produced ionic and
radical fragments as triggers of the subsequent ion-radical photochemistry, \eg, on ice mantels of dust
particles in interstellar space. They also generally help to disentangle the broad inventory of
environmentally important radicals.

\section{Data Availability}
The data that support the findings of this study are available from the corresponding author upon
request.

\section{Code Availability}
The scripts used to analyse the recorded data and the specified equations are available from the
corresponding author upon request.

\section{Supporting Information}
Experimental ion-imaging map with the deflector-electrodes grounded;
photoion-photoion coincidence map; description of a model for calculating the appearance energies;
plot of the total-momentum release for specified ions; dependence of the appearance energy on the
relative ion yields; details on thecorrection of the background in the calculations of the
H$_2{}^{18}$O$^+$/H$_3{}^{18}$O$^+$ ratio.

\section{Acknowledgments}
We acknowledge financial support by Deutsches Elektronen-Synchrotron DESY, a member of the Helmholtz
Association (HGF), also for the use of the Maxwell computational resources operated at DESY and
through the Center for Molecular Water Science (CMWS). This work was further supported by the Matter
innovation pool of the Helmholtz Association through the DataX project and through the
Helmholtz-Lund International Graduate School (HELIOS, HIRS-0018) and by the federal cluster of
excellence ``Advanced Imaging of Matter'' (AIM, EXC~2056, ID~390715994) of the Deutsche
Forschungsgemeinschaft (DFG).

\section{References}
\label{sec:References}
\bibliography{string,cmi}
\onecolumngrid
\end{document}


\title{\textit{Supporting Information for Publication:} \\ Reaction Pathways of Water Dimer Following Single Ionization}
\author{Ivo~S.~Vinkl\'{a}rek}\cfeldesy%
\author{Hubertus~Bromberger}\cfeldesy%
\author{Nidin~Vadassery}\cfeldesy%
\author{Wuwei~Jin}\cfeldesy%
\author{Jochen~Küpper}\jkemail\cfeldesy\uhhcui\uhhphys%
\author{Sebastian~Trippel}\stemail\cmiweb\cfeldesy\uhhcui%
\date{\today}%
\maketitle%

\tableofcontents
\listoffigures
\listoftables
\onecolumngrid
\vspace*{100mm}

\begin{figure*}
   \includegraphics[width=1.0\textwidth]{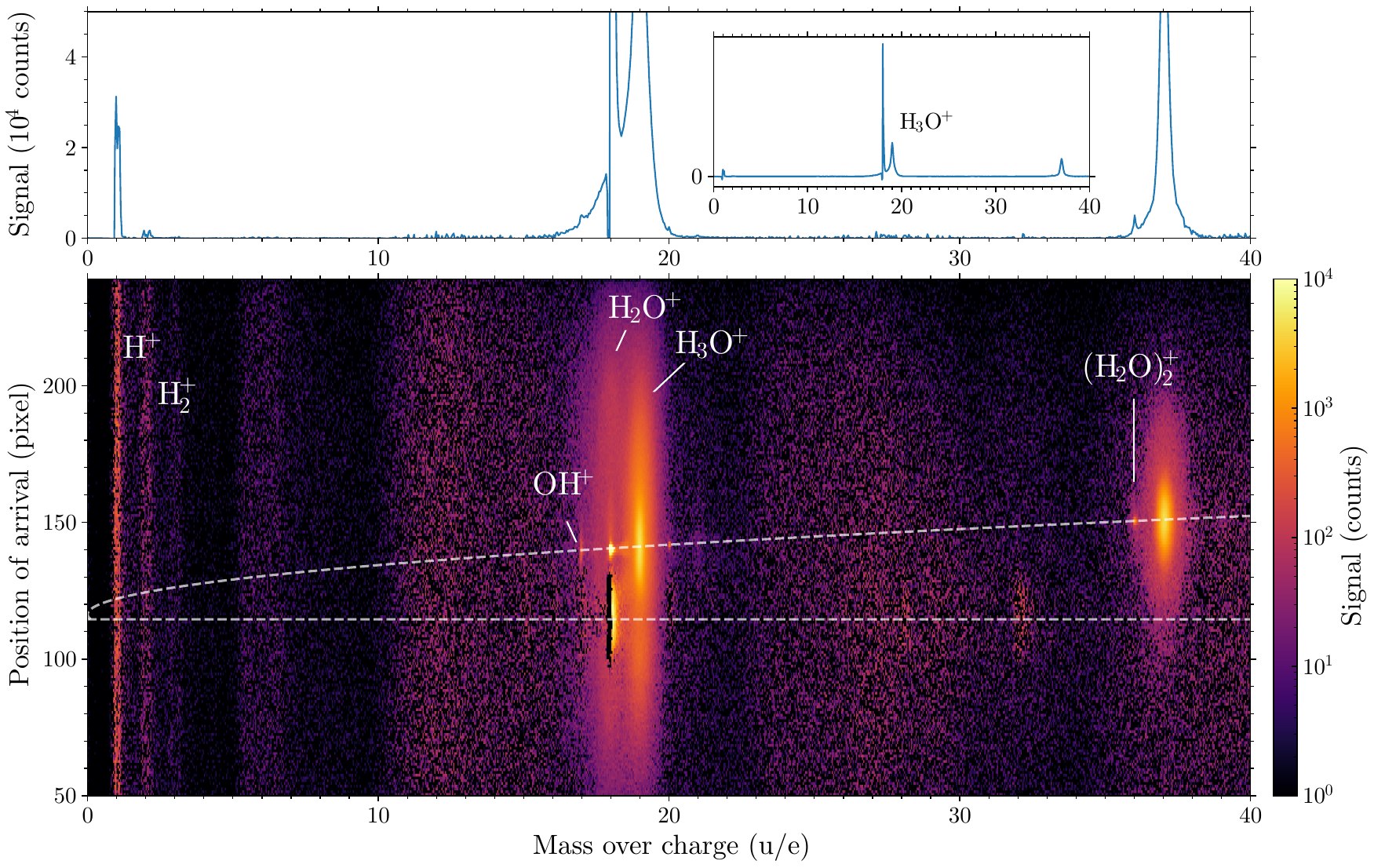}%
   \caption[Map of the ion signal without deflection]{Image mapping the ion signal according to the
      mass-over-charge ratio and the position of arrival at the detector. This two-dimensional
      spectrum was obtained with deflector off.}
   \label{fig:beamplot-non-deflected}
\end{figure*}

\pagebreak
\twocolumngrid

\section{Model}
\label{SI:sec:model}
\begin{figure}[b]
   \includegraphics[width=\linewidth]{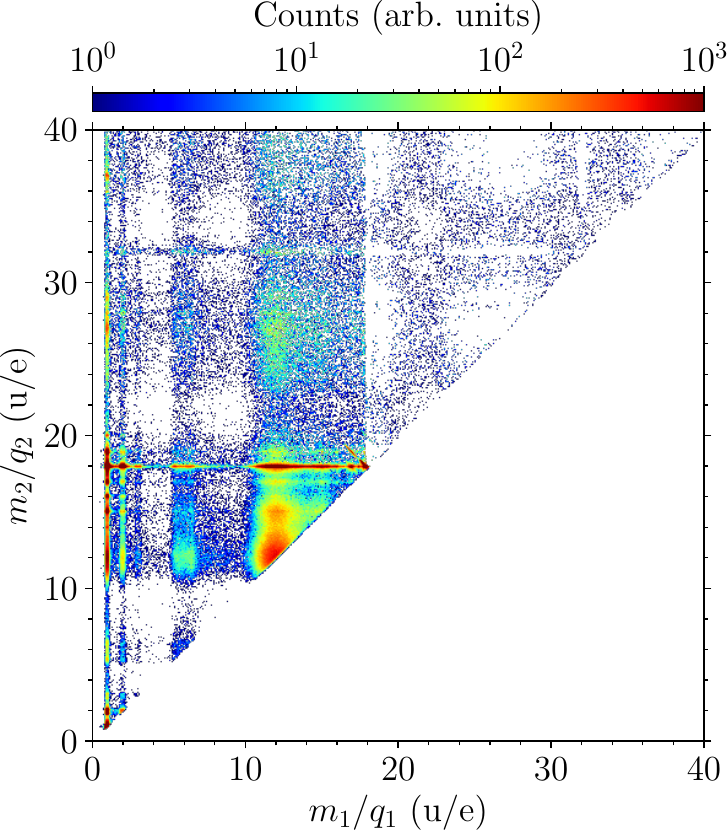}%
   \caption[Photoion-photoion correlation]{Photoion-photoion correlation for the mass-over-charge
      range between 0 and 40.}
   \label{fig:pipico0to40}
\end{figure}
\begin{table}[b]
   \begin{center}
      \caption[Branching ratios]{Comparison of the measured and calculated branching ratios for the
         selected ions.}
      \label{tab:ionyields}
      \begin{tabular}{cccccc}
        \hline
        \hline
        Ion & \HHHOp & \HHOp & \HHOdp & $\text{H}^+$ & $\text{OH}^+$ \\
        \hline
        Measured	& 0.541  & 0.176 & 0.172 & 0.060 & 0.032 \\
        Calculated 	& 0.494 & 0.180 & 0.233 & 0.061 & 0.033 \\
        \hline
        \hline
      \end{tabular}
   \end{center}
\end{table}
\begin{figure}
   \includegraphics[width=\linewidth]{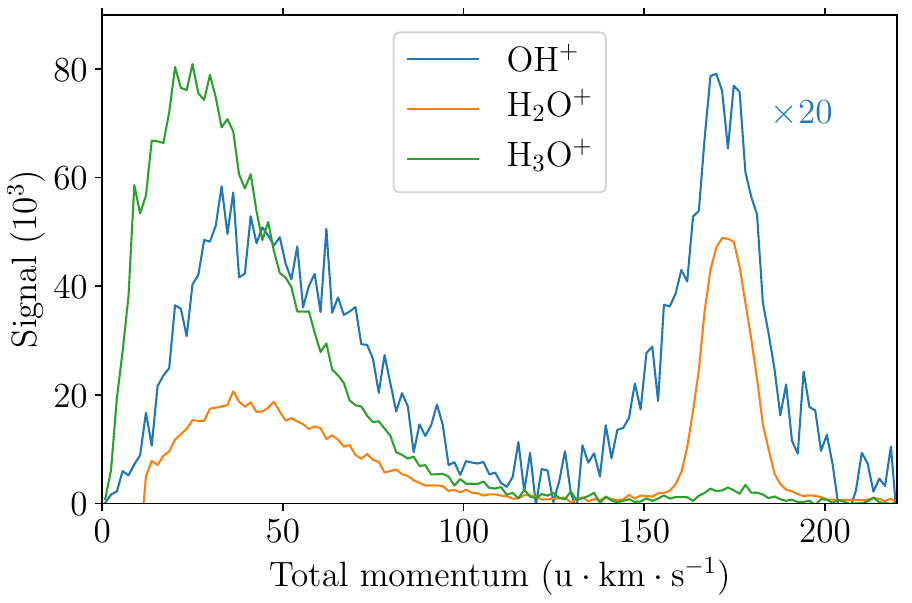}%
   \caption[Total-momentum releases]{Total-momentum release of the reaction channels yielding the
      specified ions with the assumption of two-body fragmentation.}
   \label{fig:TotalMomentum}
\end{figure}
\begin{figure}[t]
   \includegraphics[width=\linewidth]{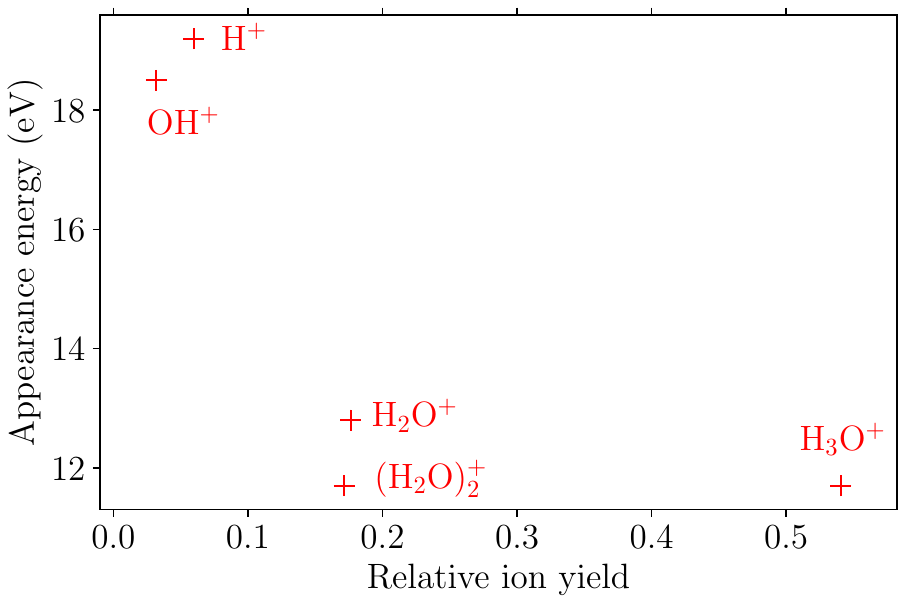}%
   \caption[Appearance energies]{Dependence of the appearance energy $E_{\text{A}}$ on the relative
      ion yields for \HHHOp, \HHOp, \HHOdp, $\text{OH}^+$, and $\text{H}^+$ ions.}
   \label{fig:FigureSI-Prob}
\end{figure}
To estimate the appearance energies for ions with low abundances, \ie, $\text{O}_2^+$,
$\text{HO}_2^+$, $(\HHO\text{-O})^+$, $\text{H}_2^+$, and $\text{H}_3^+$, we designed a
semi-empirical model combining the measured relative ion yields with reported ionization energies
\Ei and branching ratios~\cite{Svoboda:PCCP15:11531}. For the calculation of the ion yield of
$\text{I}^+$ we applied the equation
\begin{equation}
  \label{eq:ion_yield_SI}
  P(\text{I}^+) = N\cdot\sum_{k}y^{S}_{k}(\text{I}^+)\cdot D(F, \Ei(S)),
\end{equation}
where $N$ is a normalization factor, $y^{S}_{k}$ is the branching ratio of the channel $k$ after
ionization into state $S$ and $D$ is a distribution describing the ionization probability. $D$
depends on the parameter $F$ corresponding to the strength of the applied external
field~\cite{Ammosov:SVJETP64:1191} and the ionization energy $\Ei(S)$ of state $S$.
\eqref{eq:ion_yield_SI} was rewritten into the following equations using the reported
parameters~\cite{Svoboda:PCCP15:11531}
\begin{equation}
\begin{aligned}
P(\HHHOp) = N \cdot ( & D(F, 11.7)\cdot0.635 \\
                    & + D(F, 12.8)\cdot0.66 \\
                    & + D(F, 18.2)\cdot0.14 ),
\end{aligned}
\end{equation}
\begin{equation}
\begin{aligned}
P(\HHOdp) = N \cdot ( & D(F, 11.7)\cdot0.365 \\
                        & + D(F, 12.8)\cdot0.245 ),
\end{aligned}
\end{equation}
\begin{equation}
\begin{aligned}
P(\HHOp) = N \cdot ( & D(F, 12.8)\cdot0.095 \\
                        & + D(F, 18.2)\cdot0.86 \\
                        & + D(F, 18.5)\cdot(1-p_{\text{OH}^+}) \\
                        & + D(F, 19.2)\cdot(1-p_{\text{H}^+})),
\end{aligned}
\end{equation}
\begin{equation}
\begin{aligned}
P(\text{OH}^+) = N \cdot ( & D(F, 18.5) \cdot p_{\text{OH}^+}),
\end{aligned}
\end{equation}
\begin{equation}
\begin{aligned}
P(\text{H}^+) = N\cdot ( & D(F, 19.2) \cdot p_{\text{H}^+}).
\end{aligned}
\end{equation}

The parameters $F$, $p_{\text{OH}^+}$, and $p_{\text{H}^+}$ were optimized to fit the measured and
calculated relative ion yields. Assuming $D(F, \Ei) = \exp{(-\Ei/F)}$, this yielded $F = 4.34$~eV,
$p_{\text{OH}^+} = 0.37$ and $p_{\text{H}^+} = 0.82$. The calculated and measured relative ion
yields are shown in the \autoref{tab:ionyields}. The fitted field-strength parameter $F$ was then
applied to estimate appearance energies $E_{\text{A}}$ of other ions $\text{I}^+$ with low abundance
according to
\begin{equation}
   \label{eq:apperence_energy}
   E_{\text{A}}^{\text{I}^+} = - F \cdot \log ( P(\text{I}^+) )
\end{equation}
where we set branching ratio equal to $1$. The evaluated appearance energies are shown in the fourth
column of Table 1 in the main text.

Finally, we would like to point out that the fitted value of $F=4.34$~eV corresponding to the
strength of the applied external field agrees well with the ponderomotive energy $U_\text{p}$
multiplied by $1/\text{e}$, \ie,
\begin{equation}
   \label{eq:ponderomotive_energy}
   U_\text{p}=\frac{\Ei}{2 \gamma^2}\cdot\frac{1}{\text{e}} = 4.39~\text{eV},
\end{equation}
\Ei is the ionization energy of water dimer ($11.7~\text{eV}$) and $\gamma=0.7$ the corresponding
Keldysh parameter. Thus, it is tempting to assign a real physical meaning to the parameter $F$ as
the ``averaged'' ponderomotive energy. Nevertheless, further modelling of our experiment is
necessary to confirm this hypothesis.

\section{Experimental results}
\label{SI:sec:exp_res}
Here, we provide additional information on the experimental results discussed in the main text.
\autoref{fig:beamplot-non-deflected} displays the acquired background-subtracted ion signal with the
deflector off ($U_{\text{d}} = 0$~kV). The acquired signal illustrates the direct-molecular-beam
background of our measurement. \autoref{fig:pipico0to40} shows results of our photoion-photoion
covariance analysis for the mass-over-charge range between 0 and 40. The fact that there are no
cross-correlation peaks for the range $32$--$34$~u/e support the origin of the $\text{O}_2^+$,
$\text{HO}_2^+$ and $(\HHO\text{-O})^+$ ions in the $\HHOdp$ fragmentation.

\autoref{fig:TotalMomentum} shows the total-momentum release for the specified ions following
strong-field ionization of \HHOd. \autoref{fig:FigureSI-Prob} shows the correlation between the
reported appearance energies $E_{\text{A}}$ and the measured relative ion yields.

\section{Ratio of H$_2{}^{18}$O$^+$/H$_3{}^{18}$O$^+$}
\label{sec:isotopologue}
\begin{figure}
   \includegraphics[width=\linewidth]{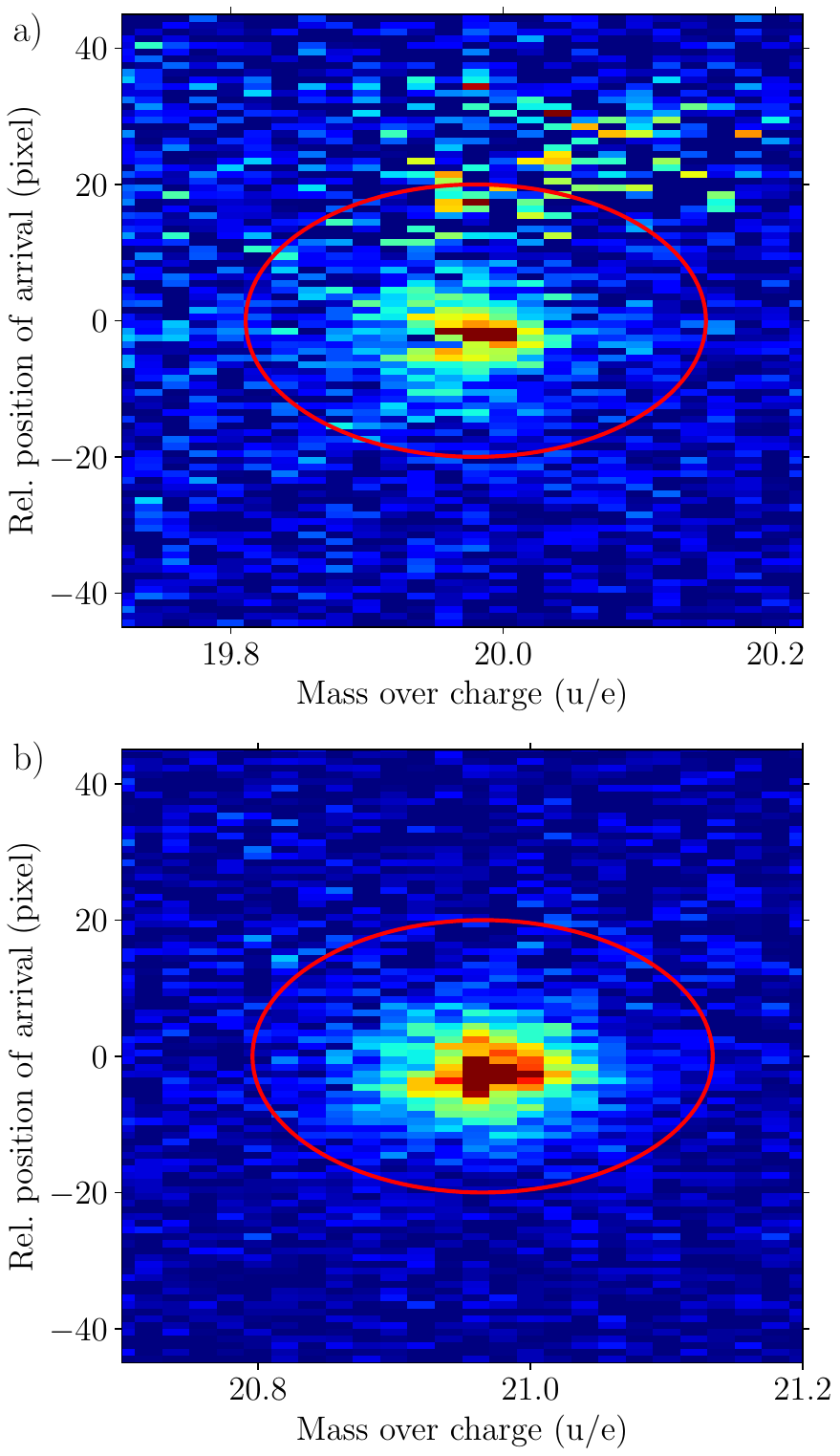}%
   \caption[Isotopologue intensities]{Ion-imaging maps for (a) H$_2{}^{18}$O$^+$ and (b) H$_3{}^{18}$O$^+$
      signals with the regions of interest used for the ion counting marked by red ellipses.}
    \label{fig:isotopologue}
\end{figure}
\begin{figure}
   \includegraphics[width=\columnwidth]{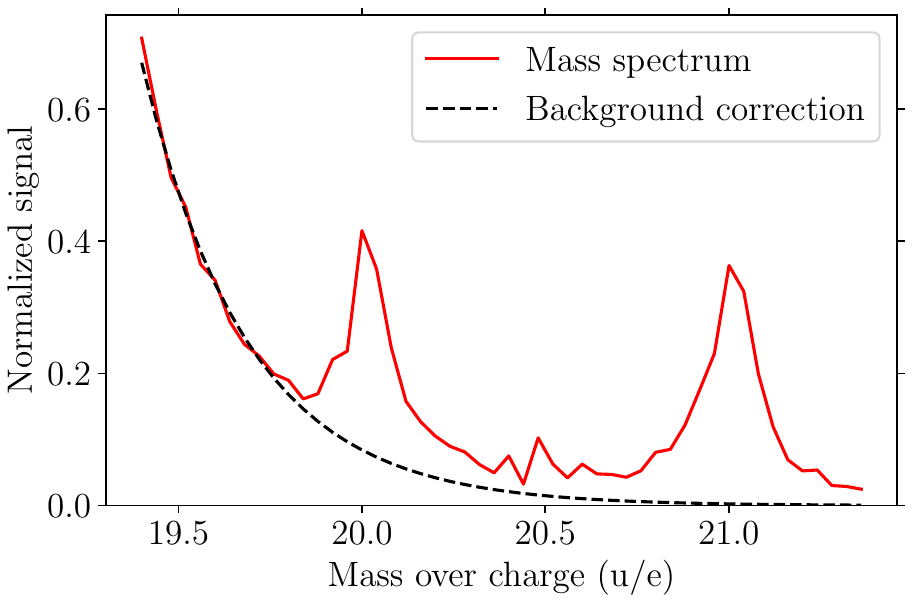}%
   \caption[Background correction]{Mass spectrum in the ${}^{18}\text{O}$-isotopologue range
      with the fitted background from ${}^{16}\text{O}$ isotopologues.}
   \label{fig:isocompMS}
\end{figure}
The ion count ratios of $\HHOp/\HHHOp\approx3$ and H$_2{}^{18}$O$^+$/H$_3{}^{18}$O$^+\approx1$,
discussed in the main text, are determined from the signals plotted in Figure~1 in the~main text.
The ratios indicate different dynamics for the respective ${}^{18}$O and ${}^{16}$O isotopologues.
\autoref{fig:isotopologue} shows the ion-imaging maps in the areas of the H$_2{}^{18}$O$^+$ and
H$_3{}^{18}$O$^+$ signals together with the region of interests used to obtain the ion counts, \ie,
10225 and 9225, respectively. This was further supported by the areas in the mass spectrum corrected
for the signal due to the broad distribution of the \HHHOp peak, see \autoref{fig:isocompMS}.

\bibliography{string,cmi}